\documentclass[amsmath,amssymb,
aps, prab, reprint]{revtex4-2}

\usepackage{graphicx}
\usepackage{dcolumn}
\usepackage{bm}
\usepackage[caption=false]{subfig}
\usepackage{siunitx}
\usepackage{tabularx}

\begin{document}
	\raggedbottom
\title{A simple approach for power density calculation of spontaneous radiation emission from a finite emittance
electron beam in planar undulators}
	
\author{Ganesh Tiwari}\email{tiwariganesh30@gmail.com}
\author{Timur Shaftan}
\affiliation{Brookhaven National Laboratory, Upton, NY, 11973, USA}

\begin{abstract}
We extend the angular power density formulae for spontaneous radiation emission from planar undulators to include finite emittance and angular misalignment of the electron beam. Then we calculate and compare power densities estimated from integral approach with Gaussian beam distribution and summation approach with the macro-particle allocation covering the particle beam phase space. After showing that both approaches converge to each other, we apply the macro-particle approach to study power absorbed and transmitted by various apertures at the front end of the 9-ID beamline at the National Synchrotron Light Source-II. Our analysis indicate that the electron beam misalignment could lead to unwarranted power deposition at tighter apertures that would have been otherwise difficult to account in the aperture design/choice process using the geometric ray-tracing approach.
\end{abstract}

\maketitle

\section{Introduction \label{sec:intro}}

Power density calculation from insertion devices in physical/angular space is necessary for determining aperture type and dimensions in the front-end of light source beamlines since it directly associates with deposited radiation dose, heat load, and activation. A rigorous approach for the power density estimation at each aperture involves propagating electric field wavefront from the source to the aperture along with obstacles in-between for each radiation frequency. Then, the total power density at the aperture can be deduced by adding the power from each frequency over the frequency range. From classical view of wave optics, this involves estimating the Fresnel and Fraunhofer diffraction integrals of radiation electric field amplitude \cite{BornWolf2005}, which becomes a non-trivial numerical problem for distances on the order of few meters and wide energy range from infrared to x-ray region (typical of undulators and wigglers in storage rings based light sources). The additive computational cost of electron trajectory calculation in insertion devices coupled with wavefront propagation over the  wide energy is quite high even with high peformance computing resources. Several available synchrotron radiation software and packages such as \texttt{SRW} \cite{srw0, srw1}, \texttt{SPECTRAX} \cite{spectra}, \texttt{ShadowOui} \cite{shadowoui}, \texttt{OSCARS} \cite{oscars}, \texttt{Sirepo} \cite{sirepo} and \texttt{OAYSIS} \cite{oasys} allow particle trajectory calculation in insertion devices and subsequent power density calculation at the desired locations using wavefront propagation or hybrid approach; still power density estimation from the electron beam passing through undulators at the aperture location integrated over all the energy space is computationally expensive with these available software because of the memory occupied by electric field wavefront structure and the extra time it takes for high precision or converging integrals. Alternate representations and wave propagation using the Wigner function approach \cite{tanaka2014} and mode decomposition to the Gauss-Hermite modes \cite{lindberg2015, tanaka2017} could subsidize the computational cost of wavefront tracking in synchrotron light sources while allowing users to retain necessary quantities such as spectral or spatial flux or brightness like expressions. The unavailability of these alternative techniques and the computational cost of wavefront propagation often results in the adoption of the geometrical ray-tracing approach for the front-end aperture design proces (see for example ~\cite{cdi, rsi}). While the geometric ray-tracing method allows us to deduce synchrotron radiation fans and limits on aperture sizes, it does not allow us to estimate the transverse distribution of radiated power density from electron beams. 

In this article, we present a simple approach to estimate the power density of spontaneously emitted radiation beams from electron beams traversing a planar undulator for applications in determining the power deposited at apertures of front-end beamlines in storage ring light sources. This approach addresses the shortcomings of geometric-ray tracing method without introducing the additional cost of wavefront propagation. We apply the solution of the paraxial wave equation for a single electron case to obtain the angular power density for a finite emittance electron beam passing through a planar undulator in section~\ref{sec:angpow}. First we apply the National Synchrotron Light Source II (NSLS-II) electron beam and 9-ID front-end beamline parameters to compare the power densities estimated using Gaussian integral and macro-particle approaches and power deposited at each aperture of the front-end for the ideal electron beam case with no misalignment. After confirming its validity with the Gaussian integral method, we adopt the macro-particle approach to estimate power absorbed by each aperture for the electron beam misalignment case in section~\ref{sec:app}. Finally, we summarize our findings in section~\ref{sec:conc}.

\section{Angular power density in planar undulators \label{sec:angpow}}

We begin with the analytical solution of paraxial wave equation. The total angular power density from a single relativistic electron can be written as \cite{Kim2017, Kim1986}

\begin{widetext}
\begin{subequations}
	\begin{align}
		\frac{d P}{d \bm{\phi}} = \frac{e^{2}}{16 \pi^{2} \epsilon_0 c T} \int d \zeta \frac{\left[\dot{\bm{\beta}}(\zeta)\dot{\bm{\tau}}(\zeta, \bm{\phi}) - \left( \dot{\bm{\beta}}(\zeta) - \bm{\phi}\right) \ddot{\bm{\tau}} (\zeta, \bm{\phi})\right]^{2}}{\dot{\bm{\tau}}^{5} (\zeta, \bm{\phi})} ,
	\end{align} 
	\text{where}
	\begin{align}
		\dot{\bm{\tau}}(\zeta, \bm{\phi}) = \frac{1}{2 \gamma^{2}} + \frac{1}{2}(\bm{\beta} - \bm{\phi})^{2}, \quad \text{and} \quad \ddot{\bm{\tau}}(\zeta, \bm{\phi}) = \dot{\bm{\beta}}.(\bm{\beta} - \bm{\phi}).
	\end{align}
	\text{Then, the numerator of the integrand takes the following form}
	\begin{align}
		\begin{split}
		&\dot{\bm{\beta}}(\zeta)\dot{\bm{\tau}}(\zeta, \bm{\phi}) - \left( \dot{\bm{\beta}}(\zeta) - \bm{\phi} \right) \ddot{\bm{\tau}} (\zeta, \bm{\phi}) \\
		&= \frac{\dot{\beta}_x}{2 \gamma^{2}} \left[ 1 - \gamma^{2} \left(\beta_x - \phi_x\right)^{2}  + \gamma^{2} \left(\beta_y - \phi_y\right)^{2} \right] \hat{x} + \dot{\beta}_x \left(\beta_x - \phi_x \right) \left(\beta_y - \phi_y\right) \hat{y}.
		\end{split}
	\end{align} 
	\label{eq:angpowden}
\end{subequations}
\end{widetext}

Here $e$ is the electron charge, $\gamma$ is its relativistic Lorentz factor, $T$ is the duration of emission, $\epsilon_0$ is vacuum permittivity, $(\beta_x, \beta_y) = (v_x/c, v_y/c)$ are electron angular velocities in $(x, y)$ and $\zeta = z/c $ for the longitudinal coordinate $z$ and speed of light $c$. $\bm{\phi} = (\phi_x, \phi_y)$ is a 2D transverse vector representing angle space. We assumed that $\dot{\beta}_y = 0$ to obtain the last expression. Since the power density is expressed in angular coordinates, we can easily infer power deposited within certain transverse area far away from the source. The advantage of this approach is that power density integration over frequency/energy domain vanishes due to delta functions thereby avoiding peculiarities of the electron energy detuning on the spectral emission as indicated in Refs.~\cite{Geloni2018, Walker2019}. However, this expression does not allow us to estimate spectral power densities, include diffraction effects and estimate power densities through multiple optical elements or apertures in physical space. 

For an electron with relativistic factor $\gamma$ passing through a planar undulator with vertical magnetic field profile given by $B_y = B_0 \text{sin}(k_u z)$ assuming negligible focusing effects and peak field strength $B_0$, the electron angular velocities are given by $(\beta_x, \beta_y) = \left(\beta_{x0} - \frac{K}{\gamma}\text{cos}(k_u z), \beta_{y0} \right)$, where $(\beta_{x0}, \beta_{y0})$ are initial angular velocities. Here the velocities follow free-space propagation and we ignore the external lattice focusing effect. We define the undulator deflection parameter $K = \frac{eB_0}{mc k_u}$ for undulator period $\lambda_u$ related to wavenumber $k_u$ by $k_u = 2 \pi/\lambda_u$; it is also defined as $K = 0.9343 \lambda_u [\text{cm}] B_0 [\text{T}]$ \cite{Kim2017}. Substituting angular velocities in Eq.~\eqref{eq:angpowden} and separating power densities polarized in $x (= \sigma)$ and $y (= \pi)$, we get
\begin{widetext}
\begin{subequations}
	\begin{align}
		\frac{dP_\sigma}{d \bm{\phi}} = \frac{e K^{2} k_u N_u}{2 \pi^{2} \epsilon_0} \frac{e \gamma^{4}}{T} \int_{0}^{2 \pi} d \xi~\text{sin}^{2}(\xi) \frac{\left( 1 - \gamma^{2} \left[\beta_{x0} - \frac{K}{\gamma} \text{cos}(\xi) - \phi_x \right]^{2}  + \gamma^{2}\left[ \beta_{y 0} - \phi_y \right]^{2} \right)^{2}}{\left( 1 + \gamma^{2} \left[\beta_{x0} - \frac{K}{\gamma} \text{cos}(\xi) - \phi_x \right]^{2}  + \gamma^{2}\left[ \beta_{y 0} - \phi_y \right]^{2} \right)^{5}}, \label{eq:sig}
	\end{align}
	\begin{align}
		\frac{dP_\pi} {d \bm{\phi} } = \frac{e K^{2} k_u N_u}{2 \pi^{2} \epsilon_0 } \frac{e \gamma^{4}}{T} \int_{0}^{2 \pi} d \xi~\text{sin}^{2}(\xi) \frac{ 4 \gamma^{2} \left( \beta_{x0} - \frac{K}{\gamma} \text{cos}(\xi) - \phi_x \right)^{2} \left( \beta_{y 0} - \phi_y \right)^{2}}{\left( 1 + \gamma^{2} \left[\beta_{x0} - \frac{K}{\gamma} \text{cos}(\xi) - \phi_x \right]^{2}  + \gamma^{2}\left[ \beta_{y 0} - \phi_y \right]^{2} \right)^{5}}. \label{eq:pi}
	\end{align}
	\label{eq:undpowden}
\end{subequations}
\end{widetext}

Here $N_u$ is total number of undulator periods. To obtain the total power density by an electron beam, we sum over contributions of each particle. Typically there are $\geq 10^{6}$ electrons making the sum calculation an extensive one. One approach to make this computation faster involves assigning pseudo-electron like or macro-particle distributions in the beam phase space with significantly less number of particles.

For a given electron beam distribution function $F(\gamma, \beta_{x0}, \beta_{y0})$, the total average power density is given by
\begin{widetext}
\begin{align}
	\left< \frac{dP}{d \bm{\phi}}\right> = \frac{1}{\int d \beta_{x0} \int d \beta_{y0} \int d \gamma~ F(\gamma, \beta_{x0}, \beta_{y0})} \int  d \beta_{x0}\int d \beta_{y0} \int d \gamma ~\frac{dP}{d \bm{\phi}} F(\gamma, \beta_{x0}, \beta_{y0}) \label{eq:avepowden}
\end{align}
In case of a normalized Gaussian electron beam distribution, the distribution takes the form
\begin{align}
	F(\gamma, \beta_{x0}, \beta_{y0}) = \frac{\text{exp}\left[-\frac{(\gamma - \overline{\gamma})^{2}}{2 \sigma_{\gamma}^{2}}\right]}{\sqrt{2 \pi} \sigma_{\gamma} } \frac{\text{exp}\left[-\frac{(\beta_{x0} - \overline{\beta}_{x0})^{2}}{2 \sigma_{\beta x}^{2}}\right]}{\sqrt{2 \pi} \sigma_{\beta x} } \frac{\text{exp}\left[-\frac{(\beta_{y0} - \overline{\beta}_{y0})^{2}}{2 \sigma_{\beta y}^{2}}\right]}{\sqrt{2 \pi} \sigma_{\beta y} }, \label{eq:normdist}
\end{align}
\end{widetext}
where $\overline{\gamma}, \overline{\beta}_{x0}, \text{ and } \overline{\beta}_{y0}$ are the corresponding beam centroids and $\sigma_{\gamma}, \sigma_{\beta x}, \text{ and } \sigma_{\beta y}$ are rms beam divergences (sizes in angular space). For this expression, the denominator in Eq.~\eqref{eq:avepowden} becomes $1$ and $e/T$ in Eq.~\eqref{eq:undpowden} should be replaced by beam current $I = N_e e /T$, where $N_e$ is the total number of electrons in the beam. The 1D integral of Eq.~\eqref{eq:undpowden} becomes a 4D integral   with non-zero beam centroids mimicing the beam offset/misalignment scenarios. This approach could be faster or slower than the  macro-particle approach depending on the required resolution for each integration in the beam phase space. For instance, if the total number of macro-particles is less than the product of total integration steps in $\gamma$, $\beta_{x0}$, and $\beta_{y0}$, the integral approach will take more time or vice-versa. 

The importance of the non-zero initial angular velocities  in Eq.~\eqref{eq:undpowden} are twofold. First, it applies to finite/non-zero emittance electron beams because they have finite size and non-zero rms divergence. Second, the tilt of the electron beam from the reference axis appears is now incorporated allowing us to estimate power densities of misaligned electron beams. To get insight into this effect quantitatively, we look at the shift in resonant fundamental wavelength and rms divergence of the radiation emitted by a single electron with non-zero initial angular velocities ($\beta_{x0}, \beta_{y0} $). Following Ref.~\cite{Kim2017}, it is easy to show that 
\begin{align}\begin{split}
&\lambda_{1}(\phi_x, \phi_y)  \\
&=\frac{\lambda_u}{2 \gamma^{2}} \left[1 + \gamma^{2} ( \beta_{x0}^{2} + \phi_{x}^{2} + \beta_{y0}^{2} + \phi_{y}^{2})  + K^{2}/2 \right], 
\end{split}
\end{align} 
where $\lambda_{1}$ is the fundamental resonant wavelength which gets shifted for non-zero initial angular velocities and angles. Since the rms divergence of this emitted photon is $\sigma_{r'} \approx \sqrt{\frac{\lambda_1}{2 L_u}}$ ~\cite{Kim2017}, the influence of the non-zero initial angular velocities is non-negligible if and only if $\beta_{x0, y0} \geq \frac{K}{\gamma \sqrt{2}}$; for electron beams, this condition becomes $\sigma_{\beta x, \beta y} \geq \frac{K}{\gamma \sqrt{2}}$. While $\sigma_{\beta x, \beta y} \ll \frac{K}{\gamma \sqrt{2}}$ implies power density approaching zero emittance case, non-zero $\beta_{x0, y0}$ in Eq.~\eqref{eq:undpowden} allow us to incorporate mean offsets of electron beam from the reference axis ($\bar{\beta}_{x0, y0} \neq 0$).

\section{ Example \label{sec:ex}}

\begin{table}[ht]%
	\caption{\label{tab:table1}%
		Parameters for angular power density calculation.} 
	\centering
		\begin{tabular}{lcc}
			\hline \hline
			\textrm{Parameter}&
			\textrm{Symbol (Unit)}&
			\textrm{Value}\\
			\hline
			\textbf{Electron beam} & \\
			Energy & $\gamma_{0}mc^{2}$ (GeV) & 3\\
			RMS energy spread & $\sigma_{\gamma}$ (MeV) & 2.67\\
			Normalized emittance in x & $\varepsilon_{nx}$ (mm. mrad) & 4.462 \\
			Normalized emittance in y & $\varepsilon_{ny}$ (mm. mrad) & 0.041 \\
			Total beam current & $I$ (Amp) & 0.5\\
			RMS divergence in x & $\sigma_{\beta x}$ ($\mu$rad) & 20.33 \\
			RMS divergence in y & $\sigma_{\beta y}$ ($\mu$rad) & 2.45 \\
			\textbf{Undulator} &\\
			Undulator period & $\lambda_{u}$ (mm) & 18\\
			Undulator length & $L_{u}$ (m) & 2.214\\
			Peak magnetic field &  $B_{0}$ (T) & 1.147\\
			\hline \hline
		\end{tabular}
\end{table}%

We consider the National Synchrotron Light Source-II (NSLS-II) electron beam and undulator parameters listed in Table \ref{tab:table1} \cite{nslsii}. The \SI{0.5}{A} current electron beam has mean energy of \SI{3}{GeV} with normalized emittances of 4.462 and \SI{0.041}{mm. \micro rad}  in horizontal and vertical directions respectively. At short straight sections, the beam have rms divergences of 20.33 and \SI{2.45}{\micro rad} in $x$ and $y$ at the undulator center. For convenience, we ignore the lattice focusing effect and consider the beam to be propagating in free-space with fixed divergences along the undulator for Eqs.~\eqref{eq:undpowden} and \eqref{eq:avepowden} to be valid. The undulator period is \SI{18}{mm} with total length of \SI{2.214}{m} and has peak magnetic field strength of \SI{1.147}{Tesla} giving rise to effective deflection factor $K=1.93$ chosen for the 9-ID of NSLS-II \cite{id}. Following discussions from the previous section, $\frac{K}{\gamma \sqrt{2}}= 2.325 \times 10^{-4}$ is much greater than rms divergences of the electron beam listed in Table~\ref{tab:table1}; therefore, the estimated radiation power density  of the finite emittance beam of Table~\ref{tab:table1} is similar to that of zero-emittance beam in the ideal case of no misalignment. 

\begin{figure}[htp]
	\centering\includegraphics[width=0.45\textwidth]{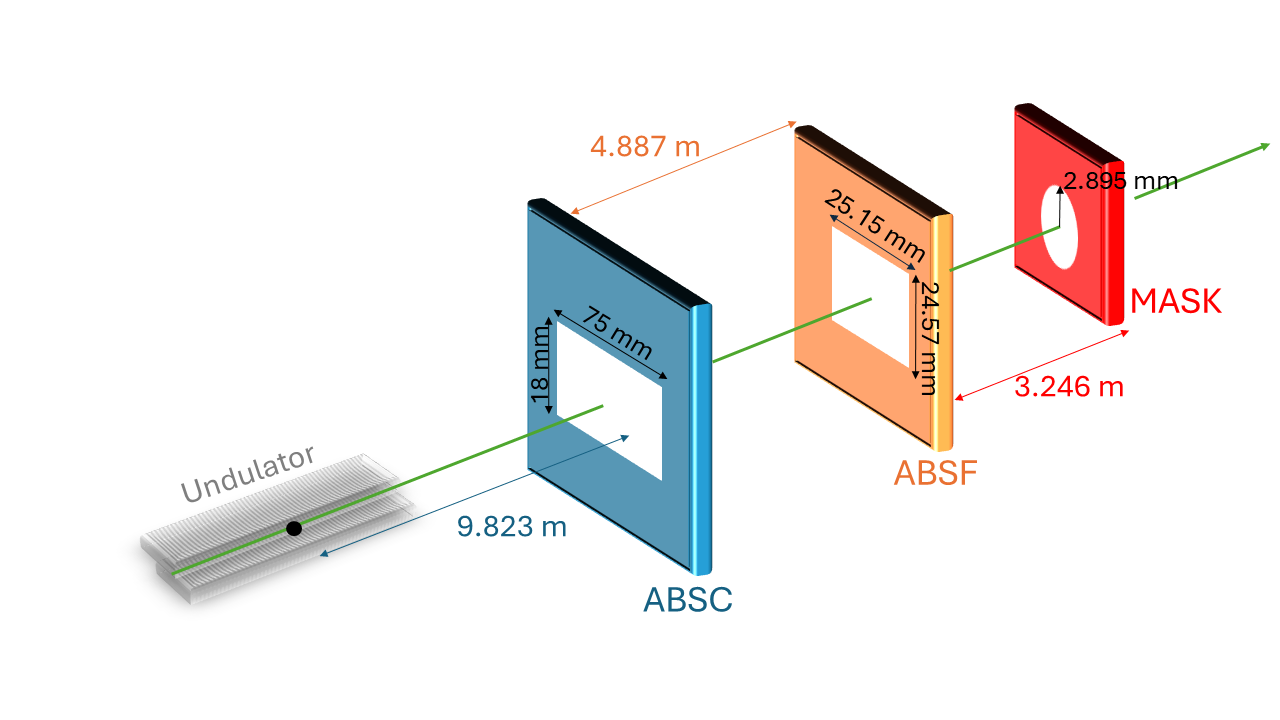}
	\caption{\label{fig:schematic} Schematic diagram of undulator and apertures with respective aperture locations, shapes and sizes. This figure is not drawn to scale.}
\end{figure} 

The schematic layout of the 9-ID front-end with relevant apertures is shown in Fig.~\ref{fig:schematic}. There are three apertures: ABSC, ABSF, and MASK; MASK is a circular aperture and the rest are of rectangular type. The first aperture, ABSC, is located at a distance of \SI{9.823}{m} from the source center and has aperture opening widths of \SI{75}{mm} in $x$ and \SI{18}{mm} in $y$; in angular space, ABSC dimensions become (7.635, 1.832) mrad. ABSF follows ABSC \SI{4.887}{m} away with aperture openings of \SI{25.15}{mm} (\SI{1.71}{mrad}) in $x$ and \SI{24.57}{m} (\SI{1.67}{mrad}) in $y$. The third and final aperture of our consideration, MASK has a diameter of \SI{5.790}{mm} (\SI{0.322}{mrad}) at a distance of \SI{17.956}{m} away from the source. In addition to these apertures, we also consider a hypthotetical circular aperture, eFWHM, with radius of \SI{0.337}{mrad} corresponding to the radiation FWHM footprint in our analyses. 

\begin{figure}[htp]
	\centering
	\begin{tabular}{c}
		\subfloat{\includegraphics[width=0.45\textwidth]{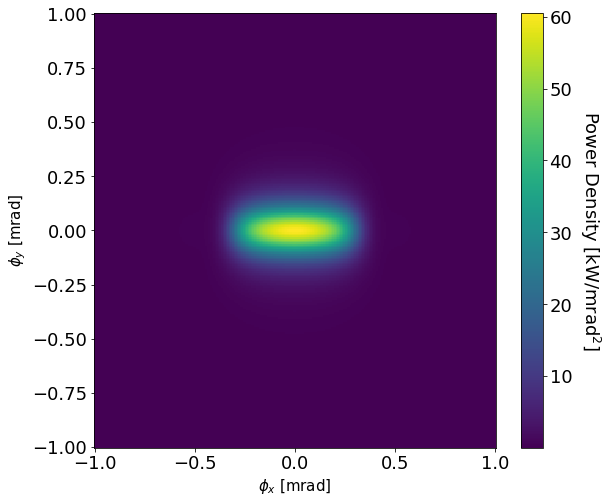}\label{subfig:powsig}}\\a\\
		\subfloat{\includegraphics[width=0.45\textwidth]{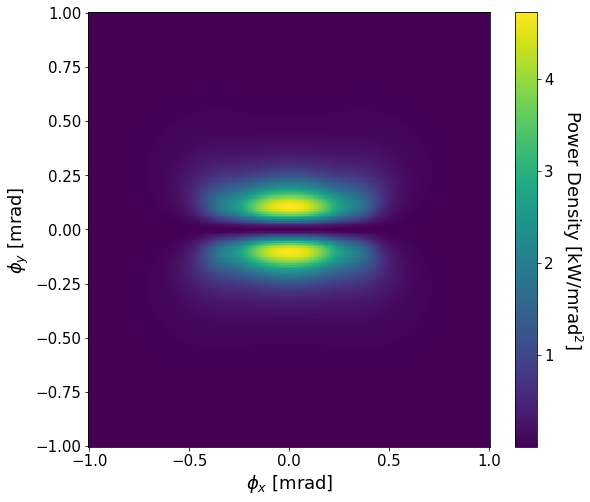}\label{subfig:powpi}}\\b\\
		\subfloat{\includegraphics[width=0.45\textwidth]{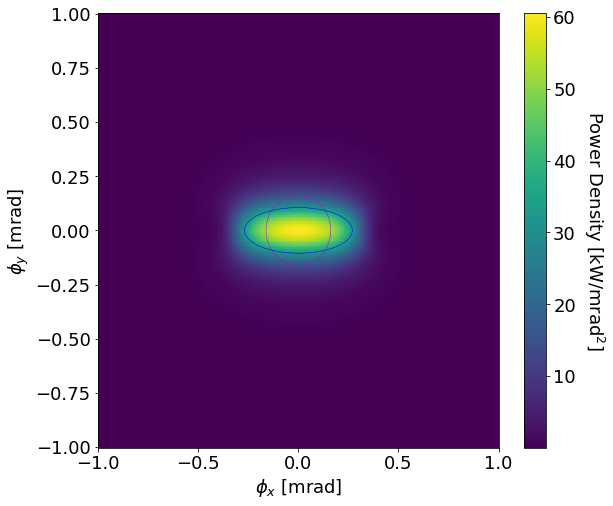}\label{subfig:powtot}}\\c
	\end{tabular}
	\caption{(a) Density plot of (a) $\sigma$ polarized, (b) $\pi$ polarized and (c) total angular power density in $\bm{\phi}$ space calculated using Eqs.~\eqref{eq:undpowden}, \eqref{eq:avepowden} and \eqref{eq:normdist}. The blue ellipse represents space occupied by FWHM values of \SI{0.54}{mrad} and \SI{0.211}{mrad} and the magenta circle has a radius of \SI{0.162}{mrad} in figure c.}\label{fig:powden}
\end{figure}

Figures \ref{subfig:powsig}-\ref{subfig:powtot} show density plot of angular power densities estimated using Eqs.~\eqref{eq:undpowden} and \eqref{eq:avepowden} for a Gaussian beam distribution of Eq.~\eqref{eq:normdist} with parameters listed in Table \ref{tab:table1} and zero mean divergence i.~e. $(\bar{\beta}_{x0}, \bar{\beta}_{y0}) = (0, 0)$. The peak power density is \SI{60.54}{kW/mrad^{2}} for $\sigma$ polarization and \SI{4.73}{kW/mrad^{2}} for $\pi$ polarization as shown in Fig.~\ref{subfig:powsig} and Fig.~\ref{subfig:powpi} respectively. The horizontal extent of both $\sigma$ and $\pi$ polarized power density in $\phi_x$ looks similar and is less than \SI{0.5}{mrad} whereas it differs by almost a factor of two in $\phi_y$ with the larger extent exhibited by $\pi$ polarization. For $\sigma$ polarized power density, the vertical width is $\sim$\SI{0.125}{mrad}. The $\pi$ polarized power density displays mirror symmetric profiles at $\phi_y =0$. We note that this double spot feature is a direct consequence of vanishing numerator near $\phi_x \sim 0$ and $\phi_y\sim 0$ in the integrand of Eq.~\eqref{eq:pi}. As a result, the $\pi$ polarized power density has vertical width of $\sim$\SI{0.25}{mrad}. However, most of the total power is contained in $\sigma$ polarization because the peak power density is almost 13 times higher than that of $\pi$ polarization. 

\begin{figure}[htpb]
	\centering\includegraphics[width=0.4\textwidth]{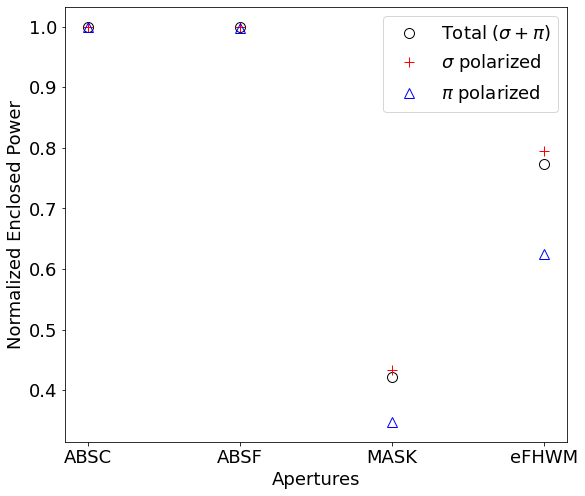}
	\caption{Normalized enclosed power calculated for $\sigma$ polarized (red $+$), $\pi$ polarized (blue $\Delta$) and total (black $\mathsf{o}$) power densities for various apertures of Fig.~\ref{fig:schematic}. ABSC and ABSF are rectangular apertures centered at the origin with respective inner dimensions of $(7.635, 1.823)$ mrad and $(1.71, 1.67)$ mrad. MASK refers to a circular aperture with radius \SI{0.1612}{mrad} whereas eFHWM has effective radius of \SI{0.3372}{mrad}. \label{fig:enpow}}
\end{figure} 

Figure~\ref{subfig:powtot} shows the density plot of total ($\sigma + \pi$)  power density in $\bm{\phi}$ space. The peak power density is \SI{60.56}{kW/mm^{2}}, roughly same as that for $\sigma$ polarized case.  We sum over power densities multiplied by the respective grid area over the whole angle space to find the total power. The total power of $\sigma$ polarized light turns out to \SI{7.21}{kW} while that of $\pi$ polarized light is \SI{1.028}{kW} giving the total power of \SI{8.238}{kW}. Using the simplified bending magnet like treatment of the undulator, we estimate the total radiated power to be $P \approx 0.63 (\gamma mc^{2})^2 [\text{GeV}] B_{0}^{2} [\text{T}]I[\text{A}] L_{u} [\text{m}] \sim$ \SI{8.26}{kW} (off by only $\sim0.27\%$ from Fig.~\ref{subfig:powtot} value of \SI{8.238}{kW}) for the given parameters in Table \ref{tab:table1} \cite{Kim2017}. The blue ellipse represents the bounded region by the full width at half maximums (FWHM) centered at the peak power density. The FWHM value in $\phi_x$ is \SI{0.54}{mrad} which is $\sim$ 26.56 times the horizontal rms divergence of the electron beam. Similarly, FWHM in $\phi_y$ is \SI{0.211}{mrad} and $\sim$ 86.12 times the vertical rms divergence of electron beam. The magenta circle represents a boundary formed by an aperture with \SI{0.322}{mrad} diameter. This corresponds to an aperture with diameter of \SI{5.79}{mm} at a distance of \SI{17.956}{m} from the undulator center. To estimate the transmitted and absorbed powers by this aperture, we calculate the normalized enclosed power using the following formulae
\begin{align}
	\begin{split}
		P_{\text{norm}}(a, b)&= 
		\frac{1}{P_{\text{total}} }\int d \phi_x \int d \phi_y ~ \frac{d P }{d \bm{\phi}}(\phi_x, \phi_y) \\
		& = \frac{2 \pi}{P_{\text{total} } } \int_{a}^{b} {r} d r ~\frac{d P}{ d \bm{\phi}} (r).
	\end{split} \label{eq:encpow}
\end{align}
Here we define angular radius $r = \sqrt{\phi_x^{2} + \phi_y^{2}}$ by adopting polar coordinates in $\bm{\phi}$ space and $P_{\text{total}}$ is the total power. This expression gives the power enclosed by an annular aperture with inner radius $a$ and outer radius $b$ from a reference center. Substituting $a=0$ allows us to estimate the power contained within  a  circular aperture of radius $b$. Figure~\ref{fig:enpow} displays the normalized enclosed power by four different apertures namely, ABSC, ABSF, MASK, and eFWHM. ABSC and ABSF have rectangular openings with dimensions of $(7.635, 1.823)$ mrad and $(1.71, 1.67)$ mrad respectively as shown in Fig.~\ref{fig:schematic}. These large apertures transmit almost $\sim$\SI{100}{\percent} of the total incident power. The power contained within the MASK of radius \SI{0.1612}{mrad} corresponds to \SI{42.2}{\percent} of the total power (\SI{43.3}{\percent} of $\sigma$-polarized and \SI{34.7}{\percent} of $\pi$-polarized powers) as indicated in Fig.~\ref{fig:enpow}. This indicates that the aperture would absorb $\sim57.8\%$ of the total power if its outer radius is greater than \SI{1.5}{mrad}. Likewise, a circular aperture with effective FWHM value $= \sqrt{\text{FWHM}_{x} \times \text{FWHM}_y} =$\SI{0.3372}{mrad} encloses $77.3\%$ of total power with almost \SI{79.5}{\percent} of $\sigma$-polarized and \SI{62.5}{\percent} of $\pi$-polarized powers as denoted by eFWHM in Fig.~\ref{fig:enpow}.

\begin{figure}[htpb]
	\centering
	\begin{tabular}{c}
		\subfloat{\includegraphics[width=0.4\textwidth]{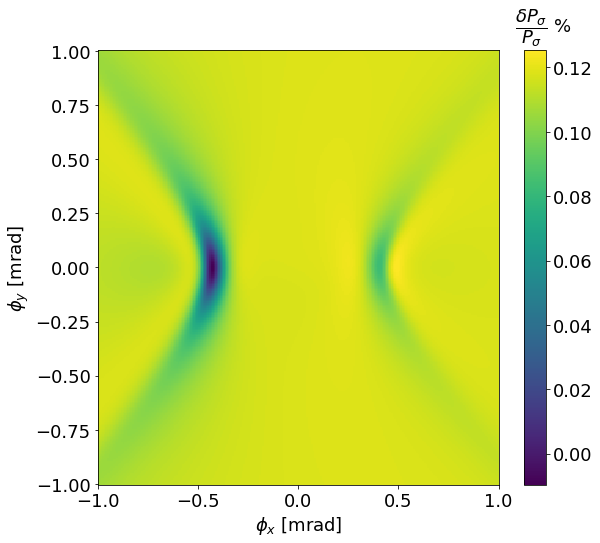}\label{subfig:diffsig}}\\(a)\\
		\subfloat{\includegraphics[width=0.4\textwidth]{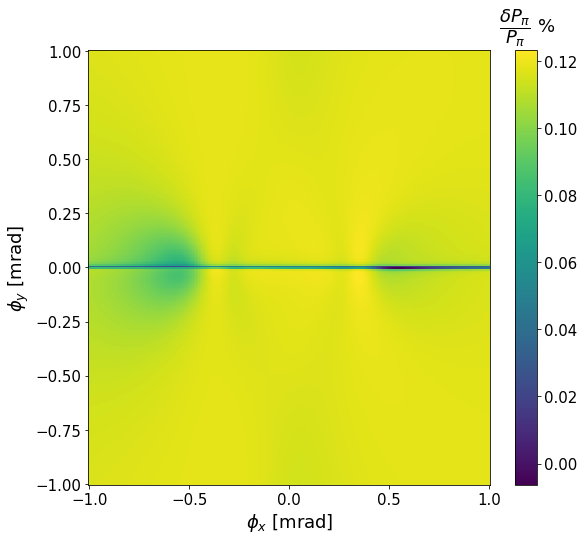}\label{subfig:diffpi}} \\
		(b) \\
	\end{tabular}
	\caption{ Density plot of normalized difference in power densities for (a) $\sigma$ and (b) $\pi$ polarization in percent. We assume the integral approach as the reference for estimating error.} \label{fig:diffpowden}
\end{figure}

\begin{figure*}[htp]
	\centering\includegraphics[width=0.9\textwidth]{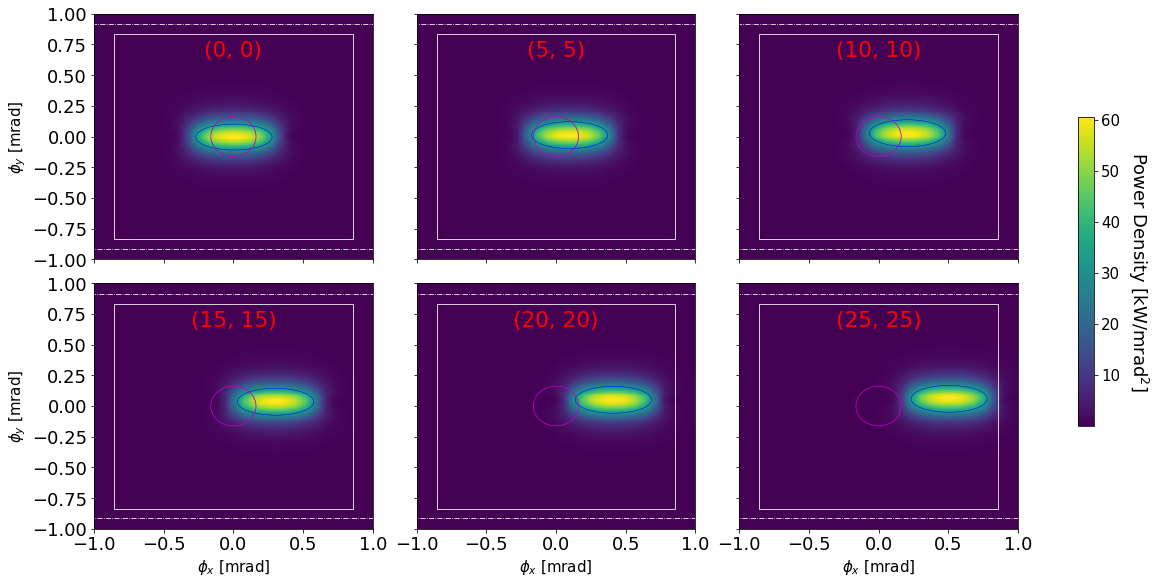}
	\caption{\label{fig:powdenall} Density plot of total power densities in angular space for misaligned electron beams with beam centriod offsets ranging from $(0, 0)$ to $(25 \sigma_{\beta x}, 25 \sigma_{\beta y})$. Blue ellipses represent area bounded by FWHM with peak densities as centers and magenta circle is a fixed mask centered at origin with \SI{0.1612}{mrad} radius corresponding to MASK for 9-ID front-end. The white dotted lines indicate boundary formed by the first rectangular aperture  ABSC and the solid white line indicates ABSF. }
\end{figure*} 

We also used the macro-particle approach using 16384 particles and Eq.~\eqref{eq:undpowden} to calculate the angular power densities. We generated the Gaussian distribution in energy and angle space by applying the Box-Muller transformation to particle allocation with the Halton sequence \cite{rpc}. Then we sum over the contributions from each particle for power density instead of evaluating three integrals: one in energy space and two in angular space. Since our previous example used 96 integration steps for each integration, the macroparticle appproach with 16384 particles would be almost 54 times faster in a single CPU processor. We compare the difference in power densities for $\sigma$ and $\pi$ polarization from the two approach. Figure \ref{fig:diffpowden} shows the normalized difference in power for (a) $\sigma$ and (b) $\pi$ polarization in percent with the integral approach values as references. The maximum percent deviation in either polarization case is $\sim$\SI{0.12}{\percent} indicating a high reliability of faster and macro-particle approach. Therefore, we adopt the multi-particle macro-particle approach for calculations in the next section. 

\section{Application\label{sec:app}}

We apply the macro-particle approach for calculating the power density in angular space to estimate the absorbed power by relevant apertures in the front-end of 9-ID of the NSLS-II \cite{cdi, rsi}. Here we consider only those apertures whose opening in one dimension is less than or equal to \SI{3}{mrad} owing to the fact that $\geq$\SI{99}{\percent} of the total emitted power is enclosed within \SI{1.5}{mrad} radius as we observed in Section \ref{sec:ex}. The schematic layout of the 9-ID front-end with relevant apertures is shown in Fig.~\ref{fig:schematic} and the radiation source parameters comprising of the electron beam and undulator are listed in Table~\ref{tab:table1}. 

\begin{table*}[ht]%
	\caption{\label{tab:table2} Power through apertures for mean offset scenarios. The effective FWHM for all cases is $\approx$\SI{0.3372}{mrad}.} 
		\begin{tabular}{lcccc}
			\hline \hline
			\textrm{Mean offset}&
			\textrm{Peak center}&  &
			\textrm{Normalized Enclosed Power by} & 
			\\
			($\bar{\beta}_{x0}/\sigma_{\beta x}, \bar{\beta}_{y0}/\sigma_{\beta y}$)&
			(x, y) [mrad] &
			ABSF [\%] &
			MASK [\%] &
			eFHWM [\%]\\
			\hline 
			(0, 0) & (0.004, 0.004)  & 100 & 42.2 & 77.3\\
			(5, 5) & (0.098, 0.012)  & 99.9 & 39.8 & 71.3\\
			(10, 10) & (0.208, 0.027) & 99.85 & 30.4 & 54.5\\
			(15, 15) & (0.302, 0.035) & 99.7 & 17.4 & 37.7\\
			(20, 20) & (0.412, 0.051) & 99.4 & 5.26 & 19.7\\
			(25, 25) & (0.506, 0.059) & 98.0 & 1.23 & 7.4\\
			\hline \hline
		\end{tabular}
\end{table*}%

Figure \ref{fig:powdenall} shows the density map of power densities in angular space for six cases of electron beam centriod beam offsets from $(0, 0)$ to $(25 \sigma_{\beta x}, 25 \sigma_{\beta y})$. We also represent the boundary region formed by colored lines or shapes. The dotted white lines represent ABSC and the solid white lines are for ABSF whereas the magenta circle represents MASK. The blue ellipses centered at peak power densities show boundary of an ellipse formed by FWHM widths. Since the FWHM values are almost fixed for all cases of misalignment, we assign a hypothetical aperture, named eFWHM with an effective radius of \SI{0.3372}{mrad}, for transmitted and absorbed power calculations. Almost all the power is transmitted by the first aperture ABSC . The normalized enclosed power values by other apertures are listed in Table \ref{tab:table2}. ABSF transmits almost \SI{100}{\percent} power for the ideal case while only absorbing about \SI{2}{\percent}, of the total incident power after ABSC, for the worst misalignment case of electron beam mean offset by 25 times  rms beam divergences in both $x$ and $y$ directions. On the other hand, MASK absorbs \SI{57.8}{\percent} of the incident power in the ideal case with the gradual increase in the absorption to \SI{98.77}{\percent} in the worst case scenario. Likewise, eFHWM absorbs significant portion of the incident power of about \SI{92.6}{\percent} in the worst beam misalignment case while blocking only \SI{22.7}{\percent} incident power for the no-misalignment case. 

\begin{figure}[htp]
	\centering\includegraphics[width=0.4\textwidth]{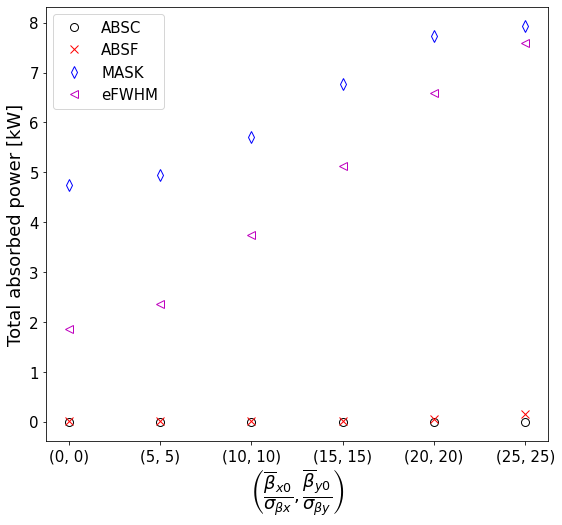}
	\caption{\label{fig:abspowall} Total absorbed power in kilowatts by each aperture for mean electron beam offsets between $(0, 0)$ and $(25 \sigma_{\beta x}, 25 \sigma_{\beta y})$.}
\end{figure}

The gradual increase in the power absorption with the beam centroid offset from ideal is best shown by Fig.~\ref{fig:abspowall} where we plot the total absorbed power (kW) versus the beam offsets for each aperture: ABSC (black circles), ABSF (red crosses), MASK (blue diamonds) and eFWHM (magenta triangles). We note that almost \SI{2}{\percent} by ABSF corresponds to $\sim$\SI{0.16}{kW}. For MASK, the absorbed power increases from \SI{4.75}{kW} at ideal case to $\sim$\SI{9.73}{kW} at the worst misalignment case. Although eFWHM starts absorbing less than half of that absorbed by MASK (\SI{1.87}{kW}), it absorbs almost same power, $\sim$\SI{7.6}{kW}, as MASK during the beam offset by $(25 \sigma_{\beta x}, 25 \sigma_{\beta y})$. This implies that the electron beam misalignment could have undesired power deposition at apertures. 

\section{Conclusion \label{sec:conc}}
To sum up, we have extended the angular power density formulae for spontaneous radiation emission from planar undulators to include finite emittance and angular misalignment of the electron beam. We have shown that the integral approach with the Gaussian beam distribution and summation approach with macro-particle allocation covering particle beam phase space  produce converging results with less than \SI{1}{\percent} deviation. Our analysis indicate that the electron beam misalignment could lead to unwarranted power deposition at smaller apertures that would been otherwise unaccounted for in the aperture design/choice process using geometric-ray tracing approach. The method and analysis presented here could be useful for aperture type and size determination for the front-ends of light source beamlines. We note that detailed power deposition calculations of finite emittance beam and misalignment scenarios along with measured/realistic magnetic field profiles of undulators is feasible through available synchrotron radiation software ~\cite{srw0, srw1, spectra, shadowoui, oscars, sirepo, oasys}. 

\begin{acknowledgments}
	This work was supported under Contract No. DE-SC0012704 with the U.S. Department of Energy. The authors thank L. Doom Jr. and M. Johanson for discussions and providing relevant documentation for 9-ID front-end and raytracing approach. We appreciate helpful remarks and suggestions from D. Hidas, O. Chubar, H. Goel, B. Nash, and D. Bruhwiler for cluster computing and on SRW capabilities. We also thank V. Smalyuk, G. Wang, G. Bassi, and Y. Li for discussions on active interlock aperture of the NSLS-II.
\end{acknowledgments}

\bibliography{empowden}

\begin{thebibliography}{20}%
\makeatletter
\providecommand \@ifxundefined [1]{%
 \@ifx{#1\undefined}
}%
\providecommand \@ifnum [1]{%
 \ifnum #1\expandafter \@firstoftwo
 \else \expandafter \@secondoftwo
 \fi
}%
\providecommand \@ifx [1]{%
 \ifx #1\expandafter \@firstoftwo
 \else \expandafter \@secondoftwo
 \fi
}%
\providecommand \natexlab [1]{#1}%
\providecommand \enquote  [1]{``#1''}%
\providecommand \bibnamefont  [1]{#1}%
\providecommand \bibfnamefont [1]{#1}%
\providecommand \citenamefont [1]{#1}%
\providecommand \href@noop [0]{\@secondoftwo}%
\providecommand \href [0]{\begingroup \@sanitize@url \@href}%
\providecommand \@href[1]{\@@startlink{#1}\@@href}%
\providecommand \@@href[1]{\endgroup#1\@@endlink}%
\providecommand \@sanitize@url [0]{\catcode `\\12\catcode `\$12\catcode
  `\&12\catcode `\#12\catcode `\^12\catcode `\_12\catcode `\%12\relax}%
\providecommand \@@startlink[1]{}%
\providecommand \@@endlink[0]{}%
\providecommand \url  [0]{\begingroup\@sanitize@url \@url }%
\providecommand \@url [1]{\endgroup\@href {#1}{\urlprefix }}%
\providecommand \urlprefix  [0]{URL }%
\providecommand \Eprint [0]{\href }%
\providecommand \doibase [0]{https://doi.org/}%
\providecommand \selectlanguage [0]{\@gobble}%
\providecommand \bibinfo  [0]{\@secondoftwo}%
\providecommand \bibfield  [0]{\@secondoftwo}%
\providecommand \translation [1]{[#1]}%
\providecommand \BibitemOpen [0]{}%
\providecommand \bibitemStop [0]{}%
\providecommand \bibitemNoStop [0]{.\EOS\space}%
\providecommand \EOS [0]{\spacefactor3000\relax}%
\providecommand \BibitemShut  [1]{\csname bibitem#1\endcsname}%
\let\auto@bib@innerbib\@empty
\bibitem [{\citenamefont {Born}\ and\ \citenamefont
  {Wolf}(2019)}]{BornWolf2005}%
  \BibitemOpen
  \bibfield  {author} {\bibinfo {author} {\bibfnamefont {M.}~\bibnamefont
  {Born}}\ and\ \bibinfo {author} {\bibfnamefont {E.}~\bibnamefont {Wolf}},\
  }\href@noop {} {\emph {\bibinfo {title} {Principles of Optics}}},\ \bibinfo
  {edition} {7th}\ ed.\ (\bibinfo  {publisher} {Cambridge University Press},\
  \bibinfo {year} {2019})\BibitemShut {NoStop}%
\bibitem [{\citenamefont {Chubar}\ and\ \citenamefont {Elleaume}(1998)}]{srw0}%
  \BibitemOpen
  \bibfield  {author} {\bibinfo {author} {\bibfnamefont {O.}~\bibnamefont
  {Chubar}}\ and\ \bibinfo {author} {\bibfnamefont {P.}~\bibnamefont
  {Elleaume}},\ }\bibfield  {title} {\bibinfo {title} {{Accurate and Efficient
  Computation of Synchrotron Radiation in the Near Field Region}},\ }in\ \href
  {https://jacow.org/e98/papers/THP01G.pdf} {\emph {\bibinfo {booktitle} {Proc.
  EPAC'98}}},\ \bibinfo {series and number} {\bibinfo {series} {European
  Particle Accelerator Conference}\ No.~\bibinfo {number} {6}}\ (\bibinfo
  {publisher} {JACoW Publishing, Geneva, Switzerland},\ \bibinfo {year}
  {1998})\ pp.\ \bibinfo {pages} {1177--1179}\BibitemShut {NoStop}%
\bibitem [{\citenamefont {Chubar}\ and\ \citenamefont {Elleaume}(2013)}]{srw1}%
  \BibitemOpen
  \bibfield  {author} {\bibinfo {author} {\bibfnamefont {O.}~\bibnamefont
  {Chubar}}\ and\ \bibinfo {author} {\bibfnamefont {P.}~\bibnamefont
  {Elleaume}},\ }\href {https://www.osti.gov//servlets/purl/1231615} {\bibinfo
  {title} {Synchrotron radiation workshop (srw), version 00}} (\bibinfo {year}
  {2013}),\ \bibinfo {note} {https://www.osti.gov/biblio/1231615}\BibitemShut
  {NoStop}%
\bibitem [{\citenamefont {Tanaka}(2021)}]{spectra}%
  \BibitemOpen
  \bibfield  {author} {\bibinfo {author} {\bibfnamefont {T.}~\bibnamefont
  {Tanaka}},\ }\bibfield  {title} {\bibinfo {title} {{Major upgrade of the
  synchrotron radiation calculation code {\it SPECTRA}}},\ }\href
  {https://doi.org/10.1107/S1600577521004100} {\bibfield  {journal} {\bibinfo
  {journal} {Journal of Synchrotron Radiation}\ }\textbf {\bibinfo {volume}
  {28}},\ \bibinfo {pages} {1267} (\bibinfo {year} {2021})}\BibitemShut
  {NoStop}%
\bibitem [{\citenamefont {Rebuffi}\ and\ \citenamefont {S{\'{a}}nchez~del
  R{\'\i}o}(2016)}]{shadowoui}%
  \BibitemOpen
  \bibfield  {author} {\bibinfo {author} {\bibfnamefont {L.}~\bibnamefont
  {Rebuffi}}\ and\ \bibinfo {author} {\bibfnamefont {M.}~\bibnamefont
  {S{\'{a}}nchez~del R{\'\i}o}},\ }\bibfield  {title} {\bibinfo {title} {{{\it
  ShadowOui}: a new visual environment for X-ray optics and synchrotron
  beamline simulations}},\ }\href {https://doi.org/10.1107/S1600577516013837}
  {\bibfield  {journal} {\bibinfo  {journal} {Journal of Synchrotron
  Radiation}\ }\textbf {\bibinfo {volume} {23}},\ \bibinfo {pages} {1357}
  (\bibinfo {year} {2016})}\BibitemShut {NoStop}%
\bibitem [{\citenamefont {Hidas}()}]{oscars}%
  \BibitemOpen
  \bibfield  {author} {\bibinfo {author} {\bibfnamefont {D.~A.}\ \bibnamefont
  {Hidas}},\ }\bibfield  {title} {\bibinfo {title} {{Novel, Fast, Open-Source
  Code for Synchrotron Radiation Computation on Arbitrary 3D Geometries}},\
  }in\ \href {https://doi.org/10.18429/JACoW-ICAP2018-TUPAG21} {\emph {\bibinfo
  {booktitle} {Proc. ICAP'18}}}\ (\bibinfo  {publisher} {JACoW Publishing,
  Geneva, Switzerland})\ pp.\ \bibinfo {pages} {309--312}\BibitemShut {NoStop}%
\bibitem [{\citenamefont {Rakitin}\ \emph {et~al.}(2018)\citenamefont
  {Rakitin}, \citenamefont {Moeller}, \citenamefont {Nagler}, \citenamefont
  {Nash}, \citenamefont {Bruhwiler}, \citenamefont {Smalyuk}, \citenamefont
  {Zhernenkov},\ and\ \citenamefont {Chubar}}]{sirepo}%
  \BibitemOpen
  \bibfield  {author} {\bibinfo {author} {\bibfnamefont {M.~S.}\ \bibnamefont
  {Rakitin}}, \bibinfo {author} {\bibfnamefont {P.}~\bibnamefont {Moeller}},
  \bibinfo {author} {\bibfnamefont {R.}~\bibnamefont {Nagler}}, \bibinfo
  {author} {\bibfnamefont {B.}~\bibnamefont {Nash}}, \bibinfo {author}
  {\bibfnamefont {D.~L.}\ \bibnamefont {Bruhwiler}}, \bibinfo {author}
  {\bibfnamefont {D.}~\bibnamefont {Smalyuk}}, \bibinfo {author} {\bibfnamefont
  {M.}~\bibnamefont {Zhernenkov}},\ and\ \bibinfo {author} {\bibfnamefont
  {O.}~\bibnamefont {Chubar}},\ }\bibfield  {title} {\bibinfo {title} {{{\it
  Sirepo}: an open-source cloud-based software interface for X-ray source and
  optics simulations}},\ }\href {https://doi.org/10.1107/S1600577518010986}
  {\bibfield  {journal} {\bibinfo  {journal} {Journal of Synchrotron
  Radiation}\ }\textbf {\bibinfo {volume} {25}},\ \bibinfo {pages} {1877}
  (\bibinfo {year} {2018})}\BibitemShut {NoStop}%
\bibitem [{\citenamefont {Rebuffi}\ and\ \citenamefont {del
  Rio}(2017)}]{oasys}%
  \BibitemOpen
  \bibfield  {author} {\bibinfo {author} {\bibfnamefont {L.}~\bibnamefont
  {Rebuffi}}\ and\ \bibinfo {author} {\bibfnamefont {M.~S.}\ \bibnamefont {del
  Rio}},\ }\bibfield  {title} {\bibinfo {title} {{OASYS (OrAnge SYnchrotron
  Suite): an open-source graphical environment for x-ray virtual
  experiments}},\ }in\ \href {https://doi.org/10.1117/12.2274263} {\emph
  {\bibinfo {booktitle} {Advances in Computational Methods for X-Ray Optics
  IV}}},\ Vol.\ \bibinfo {volume} {10388},\ \bibinfo {editor} {edited by\
  \bibinfo {editor} {\bibfnamefont {O.}~\bibnamefont {Chubar}}\ and\ \bibinfo
  {editor} {\bibfnamefont {K.}~\bibnamefont {Sawhney}}},\ \bibinfo
  {organization} {International Society for Optics and Photonics}\ (\bibinfo
  {publisher} {SPIE},\ \bibinfo {year} {2017})\ p.\ \bibinfo {pages}
  {103880S}\BibitemShut {NoStop}%
\bibitem [{\citenamefont {Tanaka}(2014)}]{tanaka2014}%
  \BibitemOpen
  \bibfield  {author} {\bibinfo {author} {\bibfnamefont {T.}~\bibnamefont
  {Tanaka}},\ }\bibfield  {title} {\bibinfo {title} {Numerical methods for
  characterization of synchrotron radiation based on the wigner function
  method},\ }\href {https://doi.org/10.1103/PhysRevSTAB.17.060702} {\bibfield
  {journal} {\bibinfo  {journal} {Phys. Rev. ST Accel. Beams}\ }\textbf
  {\bibinfo {volume} {17}},\ \bibinfo {pages} {060702} (\bibinfo {year}
  {2014})}\BibitemShut {NoStop}%
\bibitem [{\citenamefont {Lindberg}\ and\ \citenamefont
  {Kim}(2015)}]{lindberg2015}%
  \BibitemOpen
  \bibfield  {author} {\bibinfo {author} {\bibfnamefont {R.~R.}\ \bibnamefont
  {Lindberg}}\ and\ \bibinfo {author} {\bibfnamefont {K.-J.}\ \bibnamefont
  {Kim}},\ }\bibfield  {title} {\bibinfo {title} {Compact representations of
  partially coherent undulator radiation suitable for wave propagation},\
  }\href {https://doi.org/10.1103/PhysRevSTAB.18.090702} {\bibfield  {journal}
  {\bibinfo  {journal} {Phys. Rev. ST Accel. Beams}\ }\textbf {\bibinfo
  {volume} {18}},\ \bibinfo {pages} {090702} (\bibinfo {year}
  {2015})}\BibitemShut {NoStop}%
\bibitem [{\citenamefont {Tanaka}(2017)}]{tanaka2017}%
  \BibitemOpen
  \bibfield  {author} {\bibinfo {author} {\bibfnamefont {T.}~\bibnamefont
  {Tanaka}},\ }\bibfield  {title} {\bibinfo {title} {Coherent mode
  decomposition using mixed wigner functions of hermite--gaussian beams},\
  }\href {https://doi.org/10.1364/OL.42.001576} {\bibfield  {journal} {\bibinfo
   {journal} {Opt. Lett.}\ }\textbf {\bibinfo {volume} {42}},\ \bibinfo {pages}
  {1576} (\bibinfo {year} {2017})}\BibitemShut {NoStop}%
\bibitem [{\citenamefont {Bemish}\ \emph {et~al.}()\citenamefont {Bemish},
  \citenamefont {Stelmach}, \citenamefont {Johanson}, \citenamefont {Ackerman},
  \citenamefont {Cheswick},\ and\ \citenamefont {Doom}}]{cdi}%
  \BibitemOpen
  \bibfield  {author} {\bibinfo {author} {\bibfnamefont {B.}~\bibnamefont
  {Bemish}}, \bibinfo {author} {\bibfnamefont {C.}~\bibnamefont {Stelmach}},
  \bibinfo {author} {\bibfnamefont {M.}~\bibnamefont {Johanson}}, \bibinfo
  {author} {\bibfnamefont {A.}~\bibnamefont {Ackerman}}, \bibinfo {author}
  {\bibfnamefont {E.}~\bibnamefont {Cheswick}},\ and\ \bibinfo {author}
  {\bibfnamefont {L.}~\bibnamefont {Doom}},\ }\href@noop {} {\bibinfo {title}
  {{STORAGE RING FRONT END, CDI BEAMLINE RAY TRACING}}},\ \bibinfo {note}
  {{SR-FE-09ID-1101}}\BibitemShut {NoStop}%
\bibitem [{\citenamefont {Hulbert}\ \emph {et~al.}()\citenamefont {Hulbert},
  \citenamefont {Walter},\ and\ \citenamefont {Williams}}]{rsi}%
  \BibitemOpen
  \bibfield  {author} {\bibinfo {author} {\bibfnamefont {S.}~\bibnamefont
  {Hulbert}}, \bibinfo {author} {\bibfnamefont {A.}~\bibnamefont {Walter}},\
  and\ \bibinfo {author} {\bibfnamefont {G.}~\bibnamefont {Williams}},\
  }\href@noop {} {\bibinfo {title} {{RSI for the Insertion Devices and Front
  Ends for the NEXT-II Project Beamline}}},\ \bibinfo {note}
  {{NSLSII-NX2-RSI-002}}\BibitemShut {NoStop}%
\bibitem [{\citenamefont {Kim}\ \emph {et~al.}(2017)\citenamefont {Kim},
  \citenamefont {Huang},\ and\ \citenamefont {Lindberg}}]{Kim2017}%
  \BibitemOpen
  \bibfield  {author} {\bibinfo {author} {\bibfnamefont {K.-J.}\ \bibnamefont
  {Kim}}, \bibinfo {author} {\bibfnamefont {Z.}~\bibnamefont {Huang}},\ and\
  \bibinfo {author} {\bibfnamefont {R.}~\bibnamefont {Lindberg}},\ }\href
  {https://doi.org/DOI: 10.1017/9781316677377} {\emph {\bibinfo {title}
  {{Synchrotron Radiation and Free-Electron Lasers: Principles of Coherent
  X-Ray Generation}}}}\ (\bibinfo  {publisher} {Cambridge University Press},\
  \bibinfo {address} {Cambridge},\ \bibinfo {year} {2017})\BibitemShut
  {NoStop}%
\bibitem [{\citenamefont {Kim}(1986)}]{Kim1986}%
  \BibitemOpen
  \bibfield  {author} {\bibinfo {author} {\bibfnamefont {K.-J.}\ \bibnamefont
  {Kim}},\ }\bibfield  {title} {\bibinfo {title} {Angular distribution of
  undulator power for an arbitrary deflection parameter k},\ }\href
  {https://doi.org/https://doi.org/10.1016/0168-9002(86)90047-1} {\bibfield
  {journal} {\bibinfo  {journal} {Nuclear Instruments and Methods in Physics
  Research Section A: Accelerators, Spectrometers, Detectors and Associated
  Equipment}\ }\textbf {\bibinfo {volume} {246}},\ \bibinfo {pages} {67}
  (\bibinfo {year} {1986})}\BibitemShut {NoStop}%
\bibitem [{\citenamefont {Geloni}\ \emph {et~al.}(2018)\citenamefont {Geloni},
  \citenamefont {Serkez}, \citenamefont {Khubbutdinov}, \citenamefont
  {Kocharyan},\ and\ \citenamefont {Saldin}}]{Geloni2018}%
  \BibitemOpen
  \bibfield  {author} {\bibinfo {author} {\bibfnamefont {G.}~\bibnamefont
  {Geloni}}, \bibinfo {author} {\bibfnamefont {S.}~\bibnamefont {Serkez}},
  \bibinfo {author} {\bibfnamefont {R.}~\bibnamefont {Khubbutdinov}}, \bibinfo
  {author} {\bibfnamefont {V.}~\bibnamefont {Kocharyan}},\ and\ \bibinfo
  {author} {\bibfnamefont {E.}~\bibnamefont {Saldin}},\ }\bibfield  {title}
  {\bibinfo {title} {{Effects of energy spread on brightness and coherence of
  undulator sources}},\ }\href {https://doi.org/10.1107/S1600577518010330}
  {\bibfield  {journal} {\bibinfo  {journal} {Journal of Synchrotron
  Radiation}\ }\textbf {\bibinfo {volume} {25}},\ \bibinfo {pages} {1335}
  (\bibinfo {year} {2018})}\BibitemShut {NoStop}%
\bibitem [{\citenamefont {Walker}(2019)}]{Walker2019}%
  \BibitemOpen
  \bibfield  {author} {\bibinfo {author} {\bibfnamefont {R.~P.}\ \bibnamefont
  {Walker}},\ }\bibfield  {title} {\bibinfo {title} {Undulator radiation
  brightness and coherence near the diffraction limit},\ }\href
  {https://doi.org/10.1103/PhysRevAccelBeams.22.050704} {\bibfield  {journal}
  {\bibinfo  {journal} {Phys. Rev. Accel. Beams}\ }\textbf {\bibinfo {volume}
  {22}},\ \bibinfo {pages} {050704} (\bibinfo {year} {2019})}\BibitemShut
  {NoStop}%
\bibitem [{nsl()}]{nslsii}%
  \BibitemOpen
  \href@noop {} {\bibinfo {title} {{NSLS-II Accelerator Parameters}}},\
  \bibinfo {note}
  {\url{https://www.bnl.gov/nsls2/accelerator/docs/accelerator-parameters.pdf}}\BibitemShut
  {NoStop}%
\bibitem [{\citenamefont {{D. A. Hidas}}()}]{id}%
  \BibitemOpen
  \bibfield  {author} {\bibinfo {author} {\bibnamefont {{D. A. Hidas}}},\
  }\href@noop {} {}\bibinfo {note} {Private communication}\BibitemShut
  {NoStop}%
\bibitem [{\citenamefont {Lindberg}()}]{rpc}%
  \BibitemOpen
  \bibfield  {author} {\bibinfo {author} {\bibfnamefont {R.~R.}\ \bibnamefont
  {Lindberg}},\ }\href@noop {} {}\bibinfo {note} {Private
  communication}\BibitemShut {NoStop}%
\end{thebibliography}%

\end{document}